УДК 681.5.01                         DOI: 10.17587/mau.23....

**А.А. Бобцов,** доктор техн. наук, bobtsov@mail.ru,
Университет ИТМО, г. Санкт-Петербург
**В.С. Воробьев,** аспирант, v.s.vorobyev@yandex.ru,
Университет ИТМО, г. Санкт-Петербург
**Н.А. Николаев,** канд. техн. наук, доцент, nikona@yandex.ru,
Университет ИТМО, г. Санкт-Петербург
**А.А. Пыркин,** доктор техн. наук, pyrkin@itmo.ru,
Университет ИТМО, г. Санкт-Петербург
**Р. Ортега,** доктор техн. наук, romeo.ortega@itam.mx,
Instituto Tecnológico Autónomo de México, México

### Синтез адаптивного наблюдателя переменных состояния для линейного стационарного объекта при наличии шумов измерений[1]

*Статья посвящена проблеме синтеза наблюдателей переменных состояния для линейных стационарных объектов, функционирующих в условиях шумов или возмущений в канале измерения. В работе рассматривается полностью наблюдаемый линейный стационарный объект с известными параметрами. Допускается, что переменные состояния не измеряются, а измеряемая выходная переменная содержит малый по амплитуде (в общем случае по модулю меньше единицы) аддитивный шум или возмущение. Также предполагается, что относительно возмущения или шума в канале измерения не имеется никакой априорной информации (например, частотный спектр, ковариация и прочее). Хорошо известно, что для данного типа объектов получено большое количество методов синтеза наблюдателей, включая прекрасно зарекомендовавший себя на практике фильтр Калмана. При условии полной наблюдаемости и наличие некоторой априорной информации о случайном процессе (что характерно для случая, когда возмущение в канале измерения может быть представлено в виде белого шума), подходы основанные на калмановской фильтрации демонстрируют высочайшее качество сходимости оценок переменных состояния к истинным значениям. Не оспаривая многочисленные результаты, полученные с использованием применения фильтра Калмана, в данной работе рассматривается альтернативная идея построения наблюдателя переменных состояния. Альтернативность нового подхода, прежде всего связана с тем, что отпадает необходимость использования привычных подходов, базирующихся на наблюдателе Люенбергера. В работе предлагается подход, основанный на оценке неизвестных параметров (в данном случае неизвестного вектора начальных условий*

---

переменных состояния объекта) некоторой линейной регрессионной модели. В рамках предлагаемого метода, после несложного преобразования осуществляется переход от динамической системы к линейной регрессионной модели с неизвестными постоянными параметрами, содержащей шум или возмущающее воздействие. После чего предлагается новая нелинейная параметризация исходной регрессионной модели, обеспечивающая уменьшение влияния шума и синтез алгоритма идентификации неизвестных постоянных параметров с использованием процедуры динамического расширения регрессора и смешивания. В статье представлены результаты компьютерного моделирования, иллюстрирующие достижение заявленных теоретических результатов.

*Ключевые слова: адаптивный наблюдатель, линейная регрессионная модель, идентификация параметров, линейные системы*

## Введение

Задача синтеза алгоритмов оценки вектора состояния динамической системы рассматривалась многими авторами на протяжении многих десятилетий. К основополагающим работам в области построений наблюдателей можно отнести работу Давида Люенбергера [1] в которой рассмотрен вопрос синтеза наблюдателей для линейных стационарных динамических систем при отсутствии возмущений, а также работы Рудольфа Калмана (см., например [2]), в которой рассматривалась линейная динамическая система, подверженная воздействию возмущающего воздействия типа «белый шум». В настоящее время разработаны конструктивные процедуры синтеза наблюдающих устройств как полной, так и пониженной размерности для линейных стационарных динамических систем неподверженных влиянию возмущения, которые могут быть найдены в как в научной, так и учебной литературе по теории автоматического управления (см., например, [3−8]).

Несмотря на то, что проблема синтеза наблюдателей переменных состояния является хорошо изученной, она остается актуальной и в настоящее время, причем, как для нелинейных, так и для линейных объектов. Задачи синтеза наблюдателей возникают в различных областях теории и практики автоматического управления, например для линейных объектов с запаздыванием [9], системах детектирования отказов [10], нестационарных системах [11] и др.

В этой статье предлагается новый альтернативный классическим решениям подход синтеза наблюдателя вектора переменных состояния линейной стационарной системы при наличии шумов и возмущений в канале измерения. Данный подход базируется на параметрической идентификации неизвестного вектора начальных условий переменных состояния объекта управления [12, 13] и методе динамического расширения регрессора и смешивания [14−16].

## Постановка задачи

Рассматривается линейная стационарная динамическая система вида

$$\dot{\mathbf{x}} = \mathbf{A}\mathbf{x} + \mathbf{B}u, \tag{1}$$

$$y = \mathbf{C}^T\mathbf{x} + \delta. \tag{2}$$

где $\mathbf{x} \in R^n$ – неизмеряемый вектор состояния, $y \in R^1$ – выходная переменная, доступная измерению, подверженная воздействию некоторого внешнего воздействия $\delta \in R^1$ – неизвестного, ограниченного возмущения, $u \in R^1$ – сигнал управления, $\mathbf{A}$, $\mathbf{B}$ и $\mathbf{C}$, соответственно, матрица состояния и векторы входа и выхода соответствующей размерности.

*Допущение 1.* В соответствии со стандартным подходом в теории наблюдателей траектории входа и состояния системы (1), (2) полагаются ограниченными для ограниченного входа.

Требуется синтезировать наблюдатель, обеспечивающий сходимость оценок переменных состояния в область истинных значений, то есть обеспечение неравенства вида

$$\lim_{t \to \infty}(x_i - \hat{x}_i) \le \varepsilon_i, \tag{3}$$

где $\varepsilon_i > 0$ – некоторое малое число, $i = 1 \dots n$.

Хорошо известно, что для оценки состояния динамической системы вида (1), (2) при отсутствии возмущения может быть использован наблюдатель полного порядка [3−8], который основан на работе [1]. При синтезе наблюдателя рассматривается дифференциальное уравнение вида

$$\dot{\hat{\mathbf{x}}} = \mathbf{A}\hat{\mathbf{x}} + \mathbf{B}u + \mathbf{L}(y - \mathbf{C}^T\hat{\mathbf{x}}), \tag{4}$$

где $\hat{\mathbf{x}}$ – обозначает оценку вектора состояния $\mathbf{x}$, $\mathbf{L}$ – матрица обратных связей (или усиления [5]) наблюдателя.

Задача синтеза наблюдателя вида (4) может быть решена только если система полностью наблюдаема (восстанавливаема [7]). Или для объекта (1) пара $\{A, C\}$ должна быть полностью наблюдаема (восстанавливаема), что в свою очередь означает существование матрицы $L$, обеспечивающий асимптотическую сходимость $\hat{\mathbf{x}}$ к $\mathbf{x}$ (см., например, [7]).

Однако при наличии шума в измерении наиболее распространенным и оптимальным решением является фильтр Калмана-Бьюси [6, 7].

## Основной результат

В работе предлагается алгоритм оценки вектора состояния линейной непрерывной системы вида (1), (2), основанный на применении подхода к синтезу наблюдателей, базирующегося на оценке параметров (*GPEBO - generalized parameter estimation based observer* [12]) и динамического расширения и смешивания регрессора (*DREM – dynamic regressor extension and mixing* [14−16]).

В соответствии с обобщенным подходом к синтезу наблюдателей, основанном на оценке параметров [12] введем в рассмотрение динамическую систему вида

$$\dot{\boldsymbol{\xi}} = \mathbf{A}\boldsymbol{\xi} + \mathbf{B}u, \tag{5}$$

и модель ошибки

$$\mathbf{e} = \mathbf{x} - \boldsymbol{\xi}. \tag{6}$$

Для производной функции ошибки (6) имеем

$$\dot{\mathbf{e}} = \mathbf{A}\mathbf{x} + \mathbf{B}u - \mathbf{A}\boldsymbol{\xi} - \mathbf{B}u = \mathbf{A}\mathbf{e},$$

решение для уравнения ошибки находится в виде

$$\mathbf{e}(t) = \boldsymbol{\Phi}(t)\mathbf{e}(0), \quad \boldsymbol{\Phi}(0) = \mathbf{I}_{n \times n}, \tag{7}$$

где $\mathbf{I}$ – единичная матрица.

Подставляя (7) в (6) получаем

$$\mathbf{x} = \boldsymbol{\xi} + \boldsymbol{\Phi}\boldsymbol{\theta},$$

где $\boldsymbol{\theta} = \mathbf{e}(0)$. Таким образом очевидно, что для вывода оценки вектора состояния динамической системы (1) необходимо получить оценки неизвестных начальных условий $\boldsymbol{\theta} = \mathbf{e}(0)$. При наличии оценок начальных условий оценка вектора

состояния может быть записана в виде

$$\hat{\mathbf{x}} = \boldsymbol{\xi} + \boldsymbol{\Phi}\hat{\boldsymbol{\theta}}, \qquad (8)$$

где $\hat{\mathbf{x}}$ и $\hat{\boldsymbol{\theta}}$ – оценки вектора состояния и начальных условий динамической системы (1), (2) соответственно.

Умножая слева предыдущее выражение на $\mathbf{C}^T$ и подставляя (2), имеем линейное регрессионное уравнение вида

$$z = \boldsymbol{\phi}\boldsymbol{\theta} + \delta, \qquad (9)$$

где $z = y - \mathbf{C}^T\boldsymbol{\xi}$ и $\boldsymbol{\phi} = \mathbf{C}^T\boldsymbol{\Phi}$.

Таким образом задача оценки вектора состояния исходной линейной динамической системы вида (1), (2) сводится к идентификации неизвестных параметров линейного регрессионного уравнения (см., например, [17, 18]).

Для реализации алгоритма оценки неизвестных параметров $\theta$ первым шагом применим алгоритм динамического расширения регрессора и смешивания [14]…[16], в соответствии с которым к линейному регрессионному уравнению (9) применяется $n - 1$ линейных, $\mathcal{L}_\infty$ - устойчивых операторов $H_i()$, $i = 1 \dots n - 1$. В качестве таких операторов можно использовать, например, апериодическое звено или звено запаздывания.

Таким образом, после применения оператора $H_i()$, уравнение (9) может быть переписано в виде

$$\bar{\mathbf{z}} = \bar{\boldsymbol{\phi}}\boldsymbol{\theta} + \bar{\boldsymbol{\delta}},$$

где $\bar{\mathbf{z}} = col(z, z_{f_1} \dots z_{f_{n-1}})$, $\bar{\boldsymbol{\phi}}^T = [\boldsymbol{\phi} \ \boldsymbol{\phi}_{f_1} \dots \boldsymbol{\phi}_{f_{n-1}}]$ и $\bar{\boldsymbol{\delta}}^T = [\delta \ \delta_{f_1} \dots \delta_{f_{n-1}}]$.

Далее из последнего уравнения может быть получено $n$ скалярных регрессоров вида

$$\bar{\mathbf{m}} = \bar{\varphi}\boldsymbol{\theta} + \bar{\boldsymbol{\delta}}_1, \qquad (10)$$

где $\bar{\mathbf{m}} = adj(\bar{\boldsymbol{\phi}})\bar{\mathbf{z}}$, $\bar{\varphi} = \det(\bar{\boldsymbol{\phi}})$ и $\bar{\boldsymbol{\delta}}_1 = adj(\bar{\boldsymbol{\phi}})\bar{\boldsymbol{\delta}}$.

*Допущение 2.* Неизвестный вектор $\bar{\boldsymbol{\delta}}_1$ такой, что выполняются неравенства вида $\left|\bar{\delta}_{1_i}\right| < 1$, $i = 1 \dots n$.

Применим к (10) линейный фильтр вида $\frac{k}{p+k}$, где $k > 0$, тогда имеем

$$\mathbf{m} = \boldsymbol{\varphi}\boldsymbol{\theta} + \boldsymbol{\delta}_1, \qquad (11)$$

где $\mathbf{m} = \frac{k}{p+k}\overline{\mathbf{m}}$, $\boldsymbol{\varphi} = \frac{k}{p+k}\overline{\boldsymbol{\varphi}}$ и $\boldsymbol{\delta}_1 = \frac{k}{p+k}\overline{\boldsymbol{\delta}}_1$.

Рассмотрим нелинейное преобразование координат вида

$$g = e^m. \tag{12}$$

Так как уравнение (11) представляет из себя $n$ скалярных регрессоров, то рассматривается процедура идентификации каждого неизвестного параметра $\theta_i$, $i = 1 \dots n$ независимо друг от друга.

$$g_i = e^{m_i} = e^{\varphi_i\theta_i + \delta_{1_i}}. \tag{13}$$

Разложим функцию $\delta_{1_i}$ в ряд Тейлора

$$\delta_{1_i} = 1 + \delta_{1_i} + \frac{\delta_{1_i}^2}{2} + \cdots. \tag{14}$$

В соответствии с предположением относительно функции $\delta_{1_i}$ из разложения рассмотрим только первые три члена, сумма которых дает значение функции близкое к ее реальному значению.

Тогда (13) можно переписать в виде

$$g_i = \left(1 + \delta_{1_i} + \frac{\delta_{1_i}^2}{2}\right)e^{\varphi_i\theta_i}. \tag{15}$$

Рассмотрим производную от функции $g_i$ (15)

$$\dot{g}_i = \dot{\varphi}_i\theta_i g_i + \frac{\dot{\delta}_{1_i} + \delta_{1_i}\dot{\delta}_{1_i}}{1 + \delta_{1_i} + \frac{\delta_{1_i}^2}{2}}g_i.$$

Выполнив ряд преобразований для последнего выражения имеем

$$\dot{g}_i + \delta_{1_i}\dot{g}_i + \frac{\delta_{1_i}^2}{2}\dot{g}_i = \dot{\varphi}_i\theta_i g_i + \frac{\delta_{1_i}^2}{2}\dot{\varphi}_i\theta_i g_i + \dot{\delta}_{1_i}g_i + \delta_{1_i}g_i m_i. \tag{16}$$

Учитывая, что $\delta_{1_i} = m_i - \varphi_i\theta_i$ и $\dot{\delta}_{1_i} = \dot{m}_i - \dot{\varphi}_i\theta_i$, после их подстановки в (16) и проведения преобразований можно получить линейное регрессионное уравнение вида

$$\varsigma_i = \boldsymbol{\psi}_i^T\boldsymbol{\Theta}, \tag{17}$$

где $\quad \varsigma_i = \dot{g}_i + m_i\dot{g}_i + 0{,}5m_i^2\dot{g}_i - g_i m_i\dot{m}_i - g_i\dot{m}_i,\qquad \boldsymbol{\psi}_i^T = [\psi_1 \quad \psi_2 \quad \psi_3]$, $\psi_1 = \varphi_i\dot{g}_i + \varphi_i m_i\dot{g}_i + g_i\dot{\varphi}_i - \varphi_i g_i\dot{m}_i + 0{,}5m_i^2\dot{\varphi}_i g_i - \dot{\varphi}_i g_i$, $\psi_2 = -0{,}5\varphi_i^2\dot{g}_i - m_i\varphi_i\dot{\varphi}_i g_i$, $\psi_3 = 0{,}5g_i\varphi_i^2\dot{\varphi}_i$ и $\boldsymbol{\Theta} = col(\theta_i \quad \theta_i^2 \quad \theta_i^3)$. Производные функций $\varphi_i$ и $m_i$ могут быть получены из фильтров, примененных в (11).

Оценку неизвестных параметров $\theta_i$, $\theta_i^2$ и $\theta_i^3$ регрессионного уравнения (17) можно выполнить используя, например, технологию динамического расширения регрессора и смешивания [14−16].

## Результаты моделирования

В качестве примера работоспособности предложенного алгоритма рассматривается динамическая система вида (1), (2) второго порядка со следующими параметрами $\mathbf{A} = \begin{bmatrix} 0 & 1 \\ -9 & 0 \end{bmatrix}$, $\mathbf{B} = \begin{bmatrix} 0 \\ 1 \end{bmatrix}$, $\mathbf{C} = \begin{bmatrix} 5 \\ 0 \end{bmatrix}$, $u = 1(t)$, начальные условия $\mathbf{x}(0) = \begin{bmatrix} 1 \\ 2 \end{bmatrix}$, возмущающее воздействие $\delta = 0{,}3\sin(t)$.

Для сравнительного анализа произведено моделирование различных схем оценки вектора состояния: классической схемы наблюдателя полного порядка, а также наблюдателя вида (8) для различных схем оценки вектора неизвестных параметров.

Для наблюдателя полной размерности в качестве вектора обратных связей выбран вектор $\mathbf{L} = \begin{bmatrix} 2 \\ 3{,}2 \end{bmatrix}$.

В качестве начальных условий модели (5) использовался вектор $\boldsymbol{\xi}(0) = \begin{bmatrix} 0 \\ 0 \end{bmatrix}$.

В процедуре динамического расширения регрессора и смешивания, для перехода к модели (10) использовался фильтра/оператор вида $\frac{1}{p+1}$.

В качестве линейного оператора, примененного в (11) использовался фильтра вида $\frac{1}{p+1}$.

Оценка неизвестных параметров модели (10) может быть выполнена, например, на базе градиентного наблюдателя вида

$$\dot{\hat{\boldsymbol{\theta}}} = -\gamma \overline{\boldsymbol{\varphi}} \left( \overline{\boldsymbol{\varphi}} \hat{\boldsymbol{\theta}} - \overline{m} \right).$$

При моделировании данной схемы использовался коэффициент адаптации $\gamma = 1$.

Для идентификации параметров модели (17) использовалась схема динамического расширения регрессора и смешивания, в которой в качестве $\mathcal{L}_\infty$ операторов использовалось апериодические звенья $H_1(p) = \frac{2}{p+2}$ и $H_2(p) = \frac{6}{p+6}$, а также коэффициенты адаптации $\gamma = 10^{11}$ при оценке параметра $\theta_1$ и $\gamma = 10^{13}$ при оценке параметра $\theta_2$.

Результаты моделирования приведены на рисунках 1…4.

На рисунках 1 и 2 приведены результаты оценки неизвестных постоянных параметров для регрессионной модели (10) с использованием градиентного алгоритма оценки, а также результаты работы предложенного алгоритма. Как видно из рисунков при сравнимом быстродействии алгоритмов, подход, предложенный в данной работе, демонстрирует более высокую точность оценки неизвестных постоянных параметров.

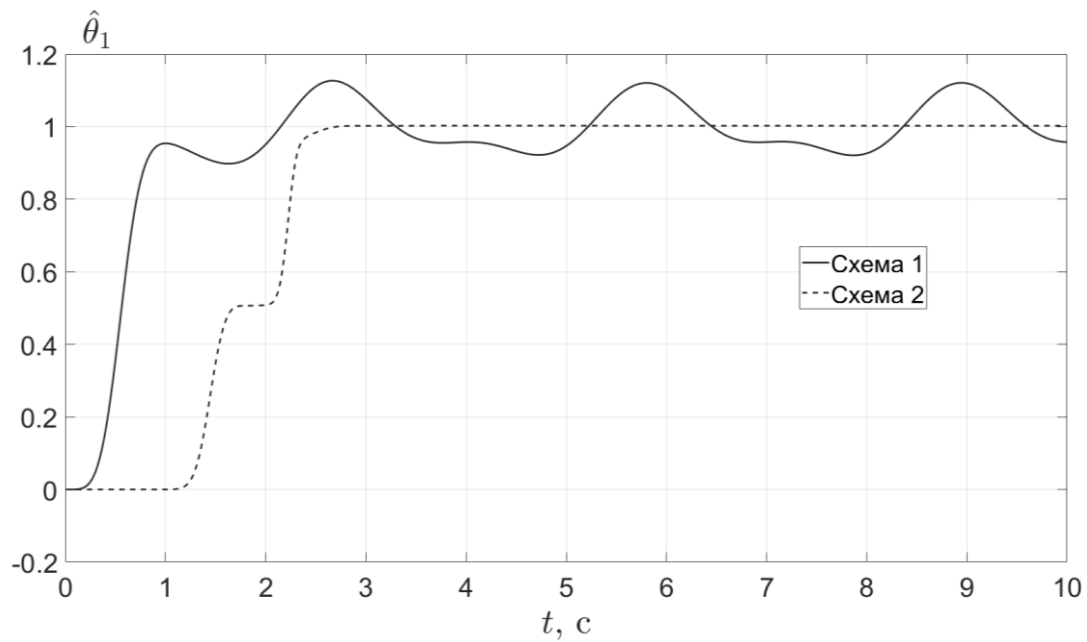

**Рис. 1.** Переходной процесс оценки неизвестного параметра $\hat{\theta}_1$, схема 1 – использование градиентного алгоритма оценки для модели (10), схема 2 – использование подхода, предложенного в данной работе

**Fig. 1.** Transients of the unknown parameter $\hat{\theta}_1$ estimate, scheme 1 – with the use of gradient observer for model (10), scheme 2 – with the use of approach proposed in this paper

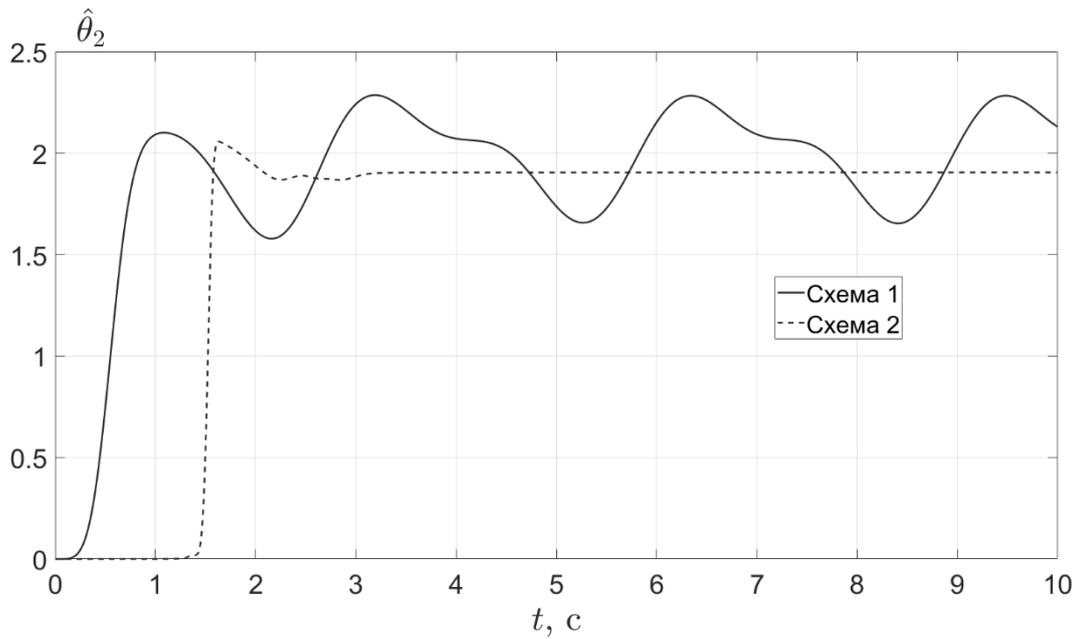

**Рис. 2.** Переходной процесс оценки неизвестного параметра $\hat{\theta}_2$, схема 1 – использование градиентного алгоритма оценки для модели (10), схема 2 – использование подхода, предложенного в данной работе

**Fig. 1.** Transients of the unknown parameter $\hat{\theta}_2$ estimate, scheme 1 – with the use of observer for model (10), scheme 2 – with the use of approach proposed in this paper

На рисунках 3 и 4 представлены результаты моделирования ошибки оценки неизвестного вектора состояния для трех алгоритмов оценки: для наблюдателя полной размерности, для алгоритма (8) в случае, когда оценка неизвестных постоянных параметров производится с использованием градиентного алгоритма оценки, примененного непосредственно к линейной регрессионной модели вида (10), а также для алгоритма (8) при использовании алгоритма оценки неизвестных постоянных параметров, предложенного в данной работе.

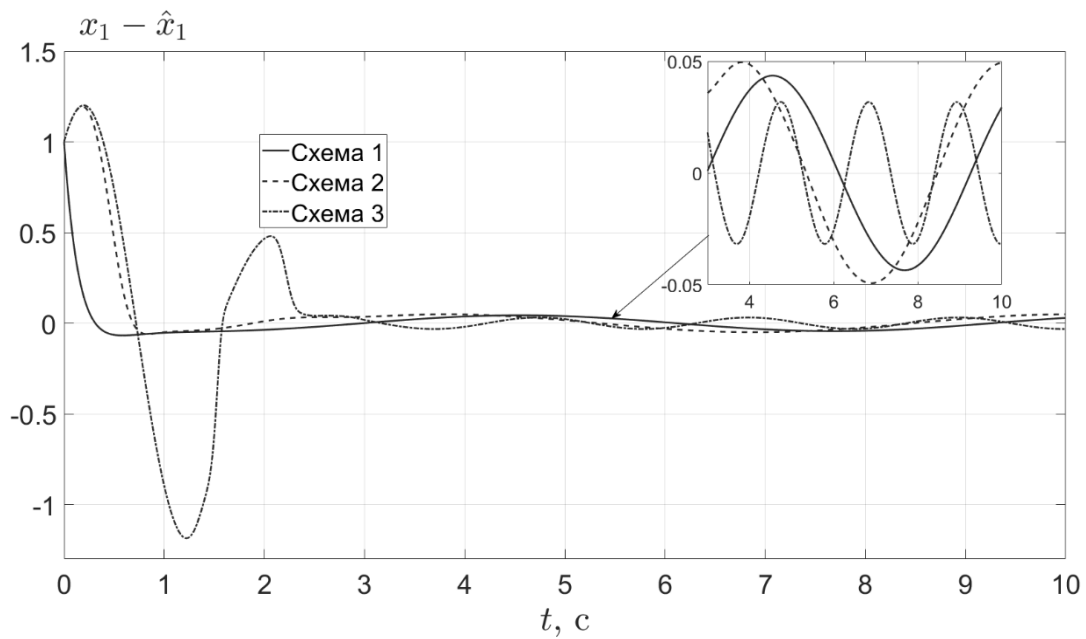

**Рис. 3.** Переходные процессы по ошибке оценивания переменной $x_1$ ($e_1 = x_1 - \hat{x}_1$) для различных схем оценки (схема 1 – оценка с помощью наблюдателя полной размерности, схема 2 – оценка с помощью уравнения (8) с идентификацией неизвестных параметров градиентным методом, схема 3 – оценка с помощью уравнения (8) с идентификацией неизвестных параметров подходом, предложенным в данной работе

**Fig. 3.** Transients of estimation error for variable $x_1$ ($e_1 = x_1 - \hat{x}_1$) for different estimation cases (scheme 1 – estimation using full order observer, scheme 2 – estimation using equation (8) and unknown parameter estimation by gradient observer, scheme 3 - estimation using equation (8) and unknown parameter estimation by approach proposed in this paper

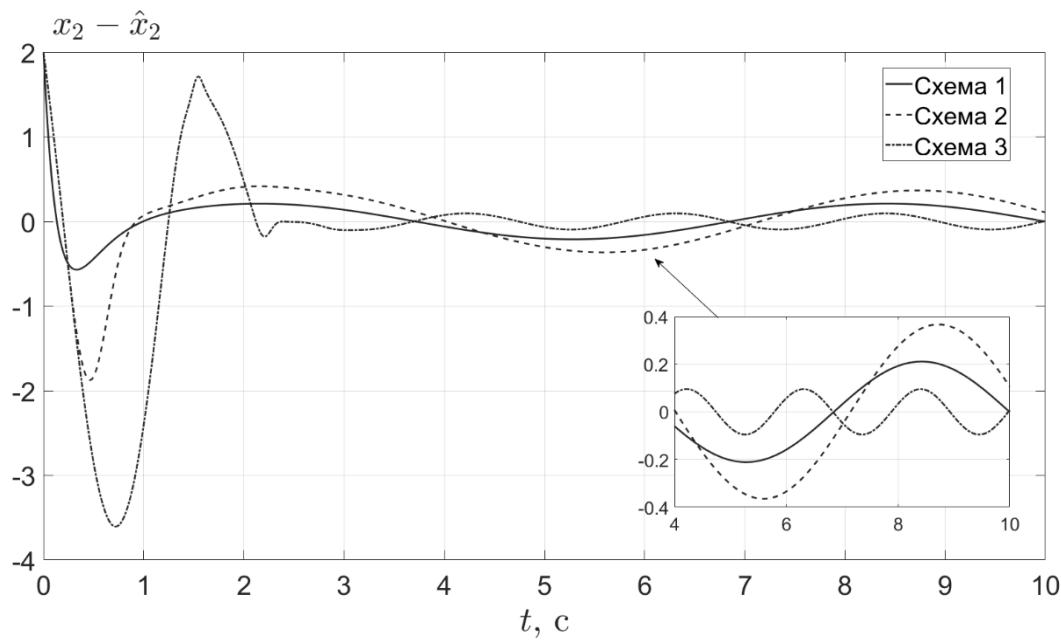

**Рис. 3.** Переходные процессы по ошибке оценивания переменной $x_2$ ($e_2 = x_2 - \hat{x}_2$) для различных схем оценки (схема 1 – оценка с помощью наблюдателя полной размерности, схема 2 – оценка с помощью уравнения (8) с идентификацией неизвестных параметров градиентным методом, схема 3 – оценка с помощью уравнения (8) с идентификацией неизвестных параметров подходом, предложенным в данной работе

**Fig. 4.** Transients of estimation error for variable $x_2$ ($e_2 = x_2 - \hat{x}_2$) for different estimation cases (scheme 1 – with the use of full order observer, scheme 2 – with the use of equation (8) and unknown parameter estimation by gradient observer, scheme 3 - with the use of equation (8) and unknown parameter estimation by approach proposed in this paper

## Заключение

В работе предложен новый подход к синтезу наблюдателя переменных состояния для линейных стационарных систем в условиях шумов измерений. Результат базируется на использовании подходов синтеза наблюдателей, основанного на оценке неизвестных параметров и динамическом расширении и смешивании регрессора. Результаты численного моделирования подтверждают работоспособность предложенного подхода.

В качестве перспектив дальнейшего развития предлагаемого подхода видится применение его к линейным нестационарным системам и объектам с неизвестными параметрами.

## Список литературы

# Synthesis of adaptive observer of state variables for a linear stationary object in the presence of measurement noise

**A.A. Bobtsov**, bobtsov@mail.ru,
ITMO University, Saint-Petersburg, 197101, Russian Federation,
**V.S.Vorobyev**, v.s.vorobyev@yandex.ru,
ITMO University, Saint-Petersburg, 197101, Russian Federation
**N.A. Nikolaev**, nikona@yandex.ru,
ITMO University, Saint-Petersburg, 197101, Russian Federation,
**A.A. Pyrkin**, pyrkin@itmo.ru,
ITMO University, Saint-Petersburg, 197101, Russian Federation,
**R. Ortega**, romeo.ortega@itam.mx,
Instituto Tecnológico Autónomo de México, 01080 México
*Corresponding author*: **A.A. Bobtsov,** Professor, ITMO University, Saint-Petersburg, 197101, Russian Federation,
e-mail: bobtsov@mail.ru

*Abstract*

*The paper is devoted to the problem of state variables observers synthesis for linear stationary system operating under condition of noise or disturbances in the measurement channel. The paper considers a completely observable linear stationary system*

with known parameters. It is assumed that the state variables are not measured, and the measured output variable contains a small amplitude (in general, modulo less than one) additive noise or disturbance. It is also assumed that there is no a priori information about the disturbance or noise in the measurement channel (for example, frequency spectrum, covariance, etc.). It is well known that many observer synthesis methods have been obtained for this type of systems, including the Kalman filter, which has proven itself in practice. Under the condition of complete observability and the presence of some a priori information about a random process (which is typical for the case when a disturbance in the measurement channel can be represented as white noise), approaches based on Kalman filtering demonstrate the highest quality estimates of state variables convergence to true values. Without disputing the numerous results obtained using the application of the Kalman filter, an alternative idea of the state variables observer constructing is considered in this paper. The alternative of the new approach is primarily due to the fact that there is no need to use the usual approaches based on the Luenberger observer. The paper proposes an approach based on the estimation of unknown parameters (in this case, an unknown vector of initial conditions of the plant state variables) of a linear regression model. Within the framework of the proposed method, after a simple transformation, a transition is made from a dynamic system to a linear regression model with unknown constant parameters containing noise or disturbing effects. After that, a new nonlinear parametrization of the original regression model and an algorithm for identifying unknown constant parameters using the procedure of dynamic expansion of the regressor and mixing are proposed which ensure reduction the influence of noise. The article presents the results of computer simulations verifying the stated theoretical results.

***Keywords****: adaptive observer, linear regression model, parameter estimation, linear systems*